\documentclass[11pt,titlepage]{article}

\usepackage{setspace}

\usepackage{amsmath}

\usepackage{graphicx}

\usepackage[round,numbers,sort&compress]{natbib}

\title{Undulatory locomotion of {\it C. elegans} on wet surfaces}

\author{Xiao~N.~Shen\thanks{Department of Mechanical Engineering and Applied Mechanics, University of Pennsylvania, Philadelphia, PA 19104, USA},
J.~Sznitman\thanks{Department of Biomedical Engineering, Technion - Israel Institute of Technology, Technion City, Haifa 32000, Israel},
P.~Krajacic\thanks{Department of Physiology, University of Pennsylvania, Philadelphia, PA 19104, USA},
T.~Lamitina\footnotemark[3],
and P.~E.~Arratia\footnotemark[1]}

\date{}

\begin{document}

\maketitle

\abstract{
The physical and bio-mechanical principles that govern undulatory movement on wet surfaces have important applications in physiology, physics, and engineering. The nematode {\it C. elegans}, with its highly stereotypical and functionally distinct sinusoidal locomotory gaits, is an excellent system in which to dissect these properties. Measurements of the main forces governing the {\it C. elegans} crawling gait on lubricated surfaces have been scarce, primarily due to difficulties in estimating the physical features at the nematode-gel interface.  Using kinematic data and a hydrodynamic model based on lubrication theory, we calculate both the surface drag forces and the nematode's bending force while crawling on the surface of agar gels.  We find that the normal and tangential surface drag force coefficients during crawling are approximately 220 and 22, respectively, and the drag coefficient ratio is approximately 10. During crawling, the calculated internal bending force is time-periodic and spatially complex, exhibiting a phase lag with respect to the nematode's body bending curvature. This phase lag is largely due to viscous drag forces, which are higher during crawling as compared to swimming in an aqueous buffer solution. The spatial patterns of bending force generated during either swimming or crawling correlate well with previously described gait-specific features of calcium signals in muscle. Further, our analysis indicates that changes in the motility gait of {\it C. elegans} is most likely due to the nematode's adaptive response to environments characterized by different drag coefficient ratios.
{\it Key words:} undulatory locomotion; lubrication; soft matter}

\clearpage

\section*{Introduction}
Many undulating organisms including earthworms and nematodes move in environments where surfaces are lubricated by a thin liquid film. Examples include soil dwelling nematodes that are responsible for soil aeration~\citep{NematodeCrawling} and parasitic nematodes that move inside the small intestine~\citep{olsen1986}. The nematode and model organism {\it Caenorhabditis elegans} is of particular interest because it is widely used in biomedical research and usually investigated while crawling on wet agar plates~\citep{Brenner,Geng, Karbowski}. Undulatory locomotion on wet surfaces bears important differences with motion on dry surfaces where the texture, structure, and patterns of the organism's outer body and surface play an important role in overcoming dry friction~\citep{RobotSnake,GraySnake, Mahadevan, Hu}. On wet surfaces, the thin liquid film formed between the organism's outer body and the surface is known to reduce (dry) friction, and hence facilitates propulsion by providing thrust. However, many questions remain regarding the forces nematodes need to generate in order to overcome surface drag. Knowledge of such forces is critical for developing accurate dynamic models~\citep{Fangyen, Cohen, Stephens}, and understanding the nematode's internal muscle activities~\citep{Pierce,Strenton,WhiteMuscle}. It can also provide insight into the different kinematics and motility gaits (e.g. swimming and crawling) observed in {\it C. elegans} studies~\citep{Cohen,Fangyen,Erdos,Pierce}.

Much effort has been devoted to understand the physical mechanisms by which organisms such as the nematode ~\textit{C. elegans} generate thrust and propulsion on lubricated surfaces~\citep{Karbowski, Gray2, Erdos, Sauvage}. Of particular interest is the drag force, and its components, acting on the surface of the nematode. Values of surface drag forces are often obtained using resistive force theory (RFT)~\citep{Cohen, Karbowski, Erdos}, born out of studies on undulatory swimming of organisms immersed in a liquid~\citep{Hancock,Brokaw, Lighthill}. In RFT, the components of the drag force normal ($F_{n}$) and tangential ($F_{t}$) to the nematode's body are assumed to be linearly proportional to the local body speed along normal and tangential directions, respectively. The corresponding drag force coefficients are $C_n$ and $C_t$. Under this scenario, the anisotropy between the normal and tangential drag coefficients (i.e. $C_{n}\neq C_{t}$) leads to net motion~\citep{LaugaPowers}. For {\it C. elegans} swimming in fluids, the drag coefficient ratio $C_{n}/C_{t}$ is typically in the range of 1.4 to 2~\citep{WuBiophyJ, Fangyen, POF}. In contrast, estimated values of $C_{n}/C_{t}$ for nematodes crawling on lubricated surfaces (agar gels) vary by as much as an order of magnitude~\citep{Cohen, Karbowski, Erdos, Sauvage}, and can range from 1.5 to 35. As a consequence, it is difficult to accurately estimate the forces and energy necessary for nematodes to move on wet surfaces.

In this study, we investigate the mechanics of undulatory locomotion on thin liquid films in experiments and using a simple model. By incorporating kinematic data into a lubrication model, we are able to estimate values of the surface drag forces and the corresponding drag coefficients of {\it C. elegans} crawling on a wet agar surface. In addition, the nematode's internal bending forces during the crawling gait are estimated by considering the contributions of internal elastic and external viscous (surface) drag forces. Finally, spatial-temporal signals of both kinematics and internal bending forces for crawling nematodes are compared to those for nematodes immersed in a fluid in order to gain further insight into the mechanisms governing these two motility gaits. The analysis provided here indicates that the adoption of a crawling gait appears to be a result of the specific anisotropy of the drag forces characterizing the wet surface rather than the high external viscous load. Results show that the spatial and temporal patterns of the internal bending force correlate well with previously described gait-specific features of calcium signals in muscle~\citep{Pierce}.

\section*{Results}

\subsection*{Experimental observations, film lubrication and anisotropic drag forces}
{\it C. elegans} are transferred and allowed to move freely on the surface of Nematode Growth Medium (NGM) agar plates, which are cultured and prepared using standard methods (see Appendix). After solidifying, a thin liquid film typically forms on top of the agar gel since the NGM water concentration is relatively high (98\% by weight). The crawling motion of the nematodes is captured using video microscopy at 15 frames per second (see SV1). A snapshot of a single nematode crawling on an agar plate over three bending (undulatory) cycles is shown in Fig. 1(a), in which the red line corresponds to the nematode's body centerline (``skeleton''). Here, the nematode's locomotion is decomposed into normal and tangential directions; the normal direction is defined to be perpendicular to the nematode body length while the tangential direction is defined along the nematode body length (Fig. 1a). The propulsive and drag forces associated with nematode crawling on thin liquid films are analyzed along these two directions.

Since the nematode's body length ($L\approx 1$mm) is approximately 25 times its body radius ($R_w\approx 40\mu$m), we consider {\it C. elegans} to be a long, slender body. Hence, the analysis (to be discussed below) is restricted to the nematode's two-dimensional cross-section (Fig. \ref{Cylinder}b). As shown in Fig. 1(a), a groove forms as the nematode moves on the agar surface~\citep{Gray2,Wallace}. In the normal direction, the nematode slides on the surface against the groove, which is lubricated by a thin liquid film. The shape of this groove is described by a symmetric circular arc curve of radius $R_a$ with an angular span of $2\theta_c$, as shown schematically in Fig. \ref{Cylinder}(c). A typical groove is 1.0 $\mu$m deep and 28 $\mu$m wide, as measured by optical interferometer~\citep{SM}. The groove effective radius and angular span are calculated to be $R_a = 100 \mu$m and $\theta_c = 20^{\circ}$. The height of the liquid film $h$ is defined as the distance between the nematode body and agar surface (Fig. \ref{Cylinder}b). For nematodes crawling on wet agar gels, this liquid film has been shown to be very thin such that $h\ll R_w$~\citep{Wallace}.

\subsection*{Analytical expressions for drag forces during crawling}
\begin{figure}[t]
\begin{center}
      \includegraphics[width=0.80\textwidth]{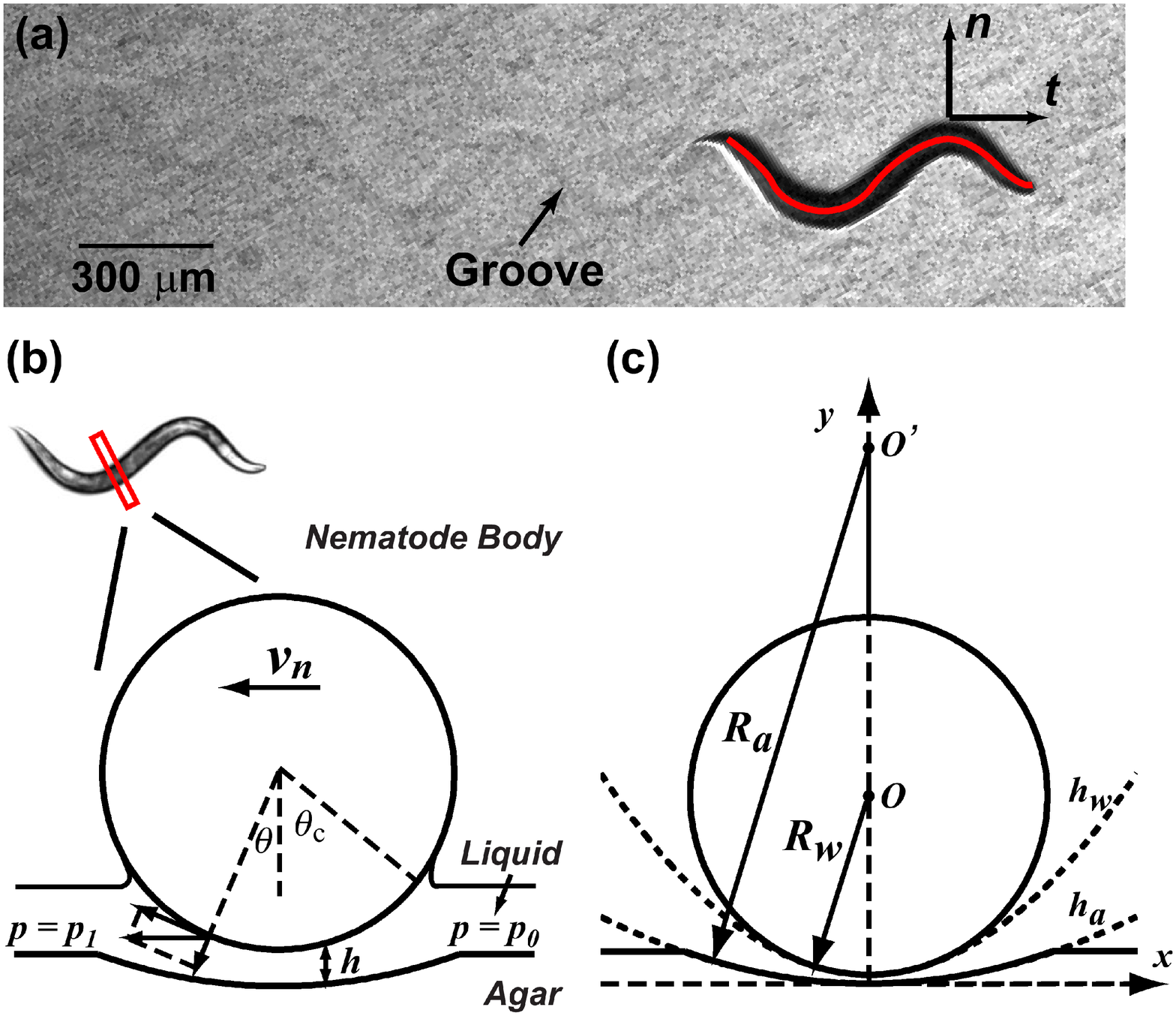}
      \caption{\label{Cylinder} (Color online) (a) The nematode {\it C. elegans} crawls on an agar plate. The red line shows the body centerline ('skeleton'). The black arrows show the unit normal and tangential vectors $\boldsymbol n$ and $\boldsymbol t$ along the nematode body. As the nematode proceed, it left over a groove on the surface. (b) Schematic of nematode cross-section during crawling. The fluid pressure on each side is $p=p_1$ and $p = p_0$, where $p_0$ is the ambient atmosphere pressure. Other parameters are defined in the text. (c) The groove shape is assumed as a circular arc with radius $R_a$. The radius of the nematode is $R_w$. Dashed lines are the parabolic curves $h_w$ and $h_a$ approximating the nematode and groove surface shape, respectively.}
\end{center}
\end{figure}

In the normal direction, the nematode's body and agar surface (i.e. groove) are approximated by two parabolic curves $h_a = x^2/2R_a$ and $h_w = h_0 + x^2/2R_w$, respectively (Fig. \ref{Cylinder}c). The liquid film thickness is $h = h_w - h_a = h_0 + (1/2) (R_w^{-1} - R_a^{-1})x^2$, where $h_0$ is the minimum film thickness. We define a groove shape parameter and a normalized film thickness as $C= (1/2) (R_w^{-1}-R_a^{-1}) R_w^2$ and $e = h_0/C$, respectively. The liquid film thickness can then be expressed as $h = C(e+X^2)$, where $X=x/R_w$. Here, the groove spans from the left edge $x=-x_m$ to the right edge $x=x_m$, with a width of $2x_m = 28~\text{}\mu$m (Fig. 1c). We note that the local curvature of the groove does not play a significant role on locomotion dynamics~\citep{Batchelor, SM}. Rather, the film thickness $h$ determines the hydrodynamics of lubrication. In other words, within the lubrication region $|x| \sim \sqrt{2R_wh}$, the flow can be effectively described by the nematode body $y=h=(h_{w}- h_{a})$ sliding on a flat surface ($y$ = $0$).

In the experiments, the nematode's typical normal sliding speed $v_n$ is $1$ mm/s, and the viscosity and density of the liquid (water) are $\mu = 1$ mPa$\cdot$s and $\rho = 10^3$ kg/m$^3$, respectively. The Reynolds number, defined as $Re = 2 \rho v_n \sqrt{2R_ah}/\mu$ is less than $10^{-2}$, and the dimensionless gap width $\delta = h/R_a \approx 0.01$ for a typical lubrication film thickness of $h \approx1 \mu$m~\citep{Hamrock}. Hence, we may omit terms in $O(\delta)$ and $O(\delta Re)$ in the dimensionless momentum equations (see detailed derivation in~\citep{SM}). If we further assume that liquid flow lubricating the nematode's body and the gel are uniform and steady, the governing equations reduce to \citep{SM}:
\begin{eqnarray}
    \frac {\partial p} {\partial y}&=& 0, \\
    \frac {\partial p} {\partial x} - \mu \frac {\partial^2 u_x} {\partial {y}^2} &=& 0, \\
    \frac {\partial u_y} {\partial y} + \frac {\partial u_x} {\partial x} &=& 0.
\end{eqnarray}
The no-slip boundary conditions on the nematode's body and the agar surface are respectively (Fig.~\ref{Cylinder}b),
\begin{flalign}
    u_x(y = 0) = 0, & \qquad {} u_x(y = h) = -v_n\cos\theta \approx -v_n, \\
    u_y(y = 0) = 0, & \qquad {} u_y(y = h) = -v_n\sin\theta = -v_n\frac {dh}{dx}.
\end{flalign}

By solving Eqs. 1-3 along with the boundary conditions, we obtain the the following velocity profiles for $u_x$ and $u_y$,
\begin{flalign}
  u_x&= \frac 1 {2\mu} \frac {d p}{d x} y^2 -\frac 1 h \left(v_n + \frac {h^2}{2\mu} \frac {d p}{d x} \right)y,\label{ux}\\
  u_y&= -\frac 1 {6\mu} \frac {d^2 p}{d x^2} y^3 + \frac 1 2 \frac {d}{d x} \left(\frac {h}{2\mu} \frac {d p}{d x} \right) y^2 - \frac {v_n} {2h^2} \frac {d h}{d x} y^2. \label{uy}
\end{flalign}
Next, we introduce the above equations for the velocity components $u_x$ and $u_y$ into Eq. 3, and obtain:
\begin{equation}
  \frac {d}{dx} \left(\frac {h^3}{12\mu} \frac {dp}{dx} \right) = -\frac {v_n} 2 \frac {dh}{dx}.
  \label{Reynolds}
\end{equation}
Equation~\ref{Reynolds} is the well known Reynolds equation~\citep{Hamrock, SM}. Note that the presence of a liquid meniscus has been observed on both sides of the nematode body as it crawls on agar gels. This liquid meniscus imparts an additional force on the nematode, namely capillary force (surface tension), in the direction of gravity. This capillary force enter in the analysis as a boundary condition (see below).

The pressure $p$ and its gradient $dp/dx$ can now be computed by rearranging and integrating Eq.~\ref{Reynolds}. The boundary conditions are (i) $d p/dx = 0$ at $\theta = \theta_c$ ($x = x_m$, $X = X_m$) and (ii) $p = p_0 + p_m$ at $\theta = \theta_c$ ($x = x_m$, $X = X_m$), where $p = p_0=0$ is the reference pressure, and $p_m$ is the capillary pressure resulting from the surface tension of liquid meniscus. The capillary pressure $p_m (\approx -125$ Pa) is estimated using the Laplace-Young equation and geometrical considerations~\citep{SM}.

This results in
\begin{eqnarray}
     \frac {dp}{dx} &=& - \frac {6\mu v_n}{C^2} \frac{X^2-X_m^2}{(e+X^2)^3}, \\
     p &=& -\frac {6\mu v_n R_i}{C^2} (J_1(e, X) - X_m^2 J_2(e,X)) + D , \label{Pressure}
\end{eqnarray}
where,
\begin{flalign*}
 &J_1(e,X) = \int_0^X t^2/(e+t^2)^3 dt, J_2(e,X) = \int_0^X 1/(e+t^2)^3 dt, \\
 &D = (6\mu v_n R_i)/C^2 (J_1(e, X_m) - X_m^2 J_2(e,X_m)) + p_m.
\end{flalign*}

Given the pressure gradient $dp/dx$, we compute the velocity profiles $u_x$ and $u_y$ in Eqs. \ref{ux} and \ref{uy}.

The drag forces acting on the nematode are obtained by integrating the fluid (water) viscous stress ($\boldsymbol \tau$) over the nematode's body surface, given by $y=h=(h_{w}- h_{a})$. Note that $\boldsymbol \tau= 2\mu\mathbf{D}$, where $\mathbf{D}$ is the rate of deformation tensor and $\mu$ is the fluid visosity. Then, the drag forces in the normal and tangential directions are given by (see derivation in \citep{SM})

\begin{flalign}
    F_x(y=h) &= \mu v_n \left\{\left(\frac {R_w}{C}\right)\left[\frac 9 2 \arctan\left(\frac {X_m}{\sqrt{e}}\right)X_m^4 e^{-\frac 5 2} + \frac 9 2 X_m^3 e^{-2} \nonumber \right.\right.\\
             &  \quad {} \left.\left. - \frac 3 2 X_m e^{-1} -\frac 1 2 \arctan\left(\frac {X_m}{\sqrt{e}}\right)e^{-\frac 1 2} \right]\right\},\label{DragForceN}
\end{flalign}
and
\begin{flalign}
     F_z(y=h) = \mu v_t \left\{\left(\frac {R_w} C\right) \left[2\arctan\left(\frac {\sin\theta_c} {\sqrt{e}}\right) e^{-1/2}\right]\right\}, \label{DragForceT}
\end{flalign}
where $v_t$ corresponds to the nematode's tangential (crawling) speed, and $F_x$ and $F_z$ are forces per unit length.

We note that the drag force relations obtained above (Eqs.~\ref{DragForceN} and \ref{DragForceT}) resemble the form of expressions obtained using resistive-force theory (RFT)~\citep{Brokaw}. Namely, the components of the drag force are linearly proportional to the fluid viscosity ($\mu$) and the nematode's crawling speed ($v_n$ and $v_t$), such that $F_{x,z}=\mu v_{n, t} C_{n, t}$, where $C_{n, t}$ is the drag coefficient and $\boldsymbol n$ and $\boldsymbol t$ correspond to the normal and tangential directions, respectively (see Fig. 1a). In Eqs.~\ref{DragForceN} and \ref{DragForceT}, the drag coefficients are defined by the expressions in the curly brackets.

The force relations derived above (Eqs.~\ref{DragForceN} and \ref{DragForceT}) can provide some useful insight. We note that as the groove surface becomes flatter, i.e. $R_a \rightarrow \infty$, the shape parameter $C= (1/2) (R_w^{-1}-R_a^{-1}) R_w^2$ increases, and the effective values of $C_n$ and $C_t$ become smaller. That is, the mere presence of a groove increases the drag forces in both normal ($x$) and tangential ($z$) directions. Also, the drag coefficient ratio $C_n/C_t$ remains independent of the groove shape parameter $C$. Therefore, according to this analysis, the presence of the groove is not a necessary condition for locomotion on agar plates, as recently suggested~\citep{Cohen}. Moreover, any change in agar stiffness affects the parameter $C$, yet it does not alter the drag coefficient ratio, in agreement with previous observations~\citep{Karbowski}.

\subsection*{Estimating the drag coefficients}
\begin{figure}
  \centering
  \includegraphics[width=0.85\textwidth]{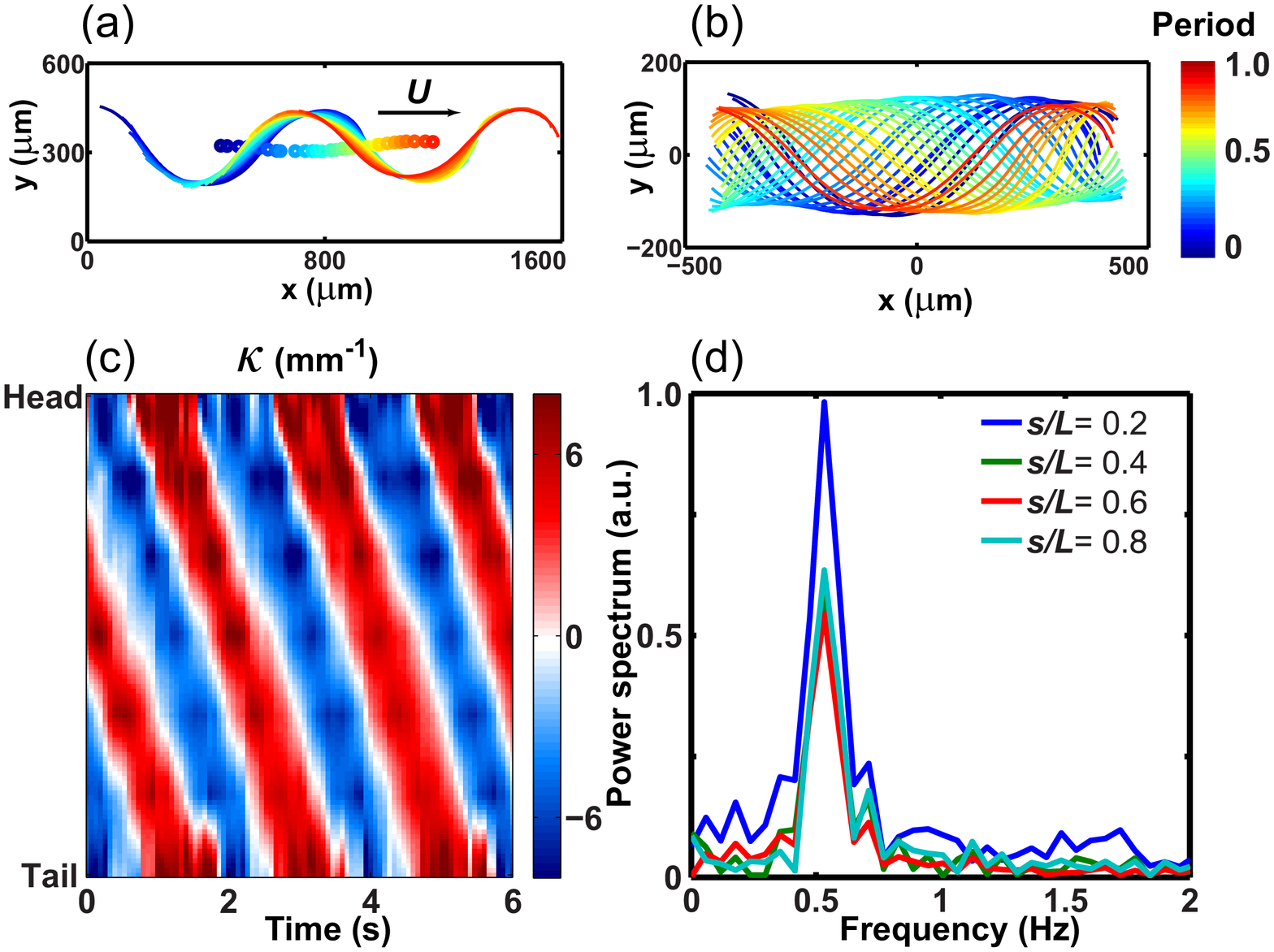}\\
  \caption{\label{UndulatoryModel}(Color online) The kinematics of {\it C. elegans} crawling on an agar plate. Body centerlines (or 'skeletons') of crawling nematodes in (a) laboratory frame and (b) nematode's frame of reference. The centroids of the skeleton in (a) are color-coded, and tracked to obtain the crawling speed $U$ (see SV2). (c) Spatio-temporal contour plot of the nematode's bending curvature $\kappa$, which display a traveling wave from head to tail. (d) Fast-Fourier transform of the curvature plot at varied locations of the nematode's body. The nematode's bending frequency is approximately 0.5 Hz and independent of position along the body.}
\end{figure}

The next step in the analysis is to calculate the drag coefficient ratio $C_n/C_t$ and its individual components for crawling motion. We begin by noting that, in resistive force theory, the ratio $C_n / C_t$ is related to the nematode's motility kinematics such that its normalized forward speed $U/c$ is given by~\cite{Gray,Karbowski, POF}:

\begin{equation}
    U/c = \frac {(2\pi^2 A^2/\lambda^2)(C_n/C_t - 1)}{(2\pi^2 A^2/\lambda^2)(C_n/C_t) + 1}.
    \label{EfficiencyEq}
\end{equation}
The quantities $c$, $A$, and $\lambda$ correspond to the nematode's wave speed, bending amplitude and wavelength, respectively. These kinematic quantities are measured in experiments by tracking the nematode's motion on agar plates as shown in Fig. 2. The nematode's motion and its kinematics are investigated using an in-house code~\cite{Plosone}. The body centroid and the body centerline (``skeleton'') are tracked and presented in Fig.~\ref{UndulatoryModel}(a) and Fig.~\ref{UndulatoryModel}(b) in the laboratory and nematode's frames, respectively. One can extract relevant kinematic metrics from the body centroids and centerlines such as forward locomotion speed $U$ and body curvature $\kappa$ (Fig. \ref{UndulatoryModel}a, c).

The curvature $\kappa$ is defined as $\kappa(s,t)=d\phi/ds$, where $\phi$ is the angle made by the tangent to the $x$-axis in the laboratory frame at each point along the body centerline, and $s$ is the arc length coordinate spanning the head of the nematode ($s=0$) to its tail ($s=L$). Body curvature is computed from the measured nematode's body centerlines $y(s,t)$, where the curvature $\kappa(s,t)=\partial^2 y / \partial x^2$ and $s \approx x$ for small amplitudes. Fig. \ref{UndulatoryModel}(c) shows the experimentally measured spatio-temporal evolution of the nematode's body curvature $\kappa(s,t)$ for three beating cycles. The periodic diagonal patterns seen in Fig.~\ref{UndulatoryModel}(b) show undulatory bending waves propagating from head to tail. The undulating waveforms are characterized by the interval of the periodic patterns and the slope of the diagonal lines. The body bending frequency ({\it f}) is obtained from the one-dimensional fast-Fourier transform of the curvature field $\kappa$ at multiple body positions $s/L$ (Fig. 2d). A single frequency peak $f\approx0.5$~Hz is found in the Fourier spectrum. This single peak is irrespective of body position and corresponds to a wave speed $c=0.39$~mm/s. The wavelength of the undulatory wave is defined as $\lambda = c/f=0.79$~mm. A summary of the experimental results is presented in Table~1 along with a comparison to data for nematodes swimming in buffer solution.

By inserting the experimentally measured values of $U$, $\lambda$, $A$ and $c$ into Eq.~\ref{EfficiencyEq}, we estimate the value of the drag coefficient ratio for crawling nematodes on wet agar surfaces to be approximately $C_{n}/C_{t} = 9.4 \pm 0.6$. This value is much larger than for nematodes immersed in Newtonian fluids ($C_{n}/C_{t}\approx2$)~\cite{Fangyen,Hancock,POF} and close to the range of values ($C_{n}/C_{t}\approx 9-14$) found for nematodes crawling on substrates of varying stiffnesses \cite{Karbowski}.

The individual components of the drag force can be calculated by evaluating the expressions enclosed by the curly brackets in Eqs.~\ref{DragForceN} and \ref{DragForceT}. These expressions are a function of the normalized span length $X = x/R_{w}$ and the normalized gap width $e = h_0/C$, where $C= (1/2) (R_w^{-1}-R_a^{-1}) R_w^2$. Here, $R_{w}=40 \mu$m and $R_{a}=100 \mu$m are the nematode's body radius and the groove's effective radius, respectively. The maximum value of span length $x_m$ is $14\text{ }\mu$m and $h_{0}=0.85\text{ }\mu$m. Hence, the values of the normal and tangential drag coefficients yield $C_n = 222.0$ and $C_t = 22.1$. To the best of our knowledge, the only other estimates available for $C_n$ and $C_t$ lie in the range of $C_n \approx C_t \approx 5-40$, with $C_n/C_t<2$ \cite{Sauvage}.

\begin{figure}
    \begin{center}
      \includegraphics[width=0.85\textwidth]{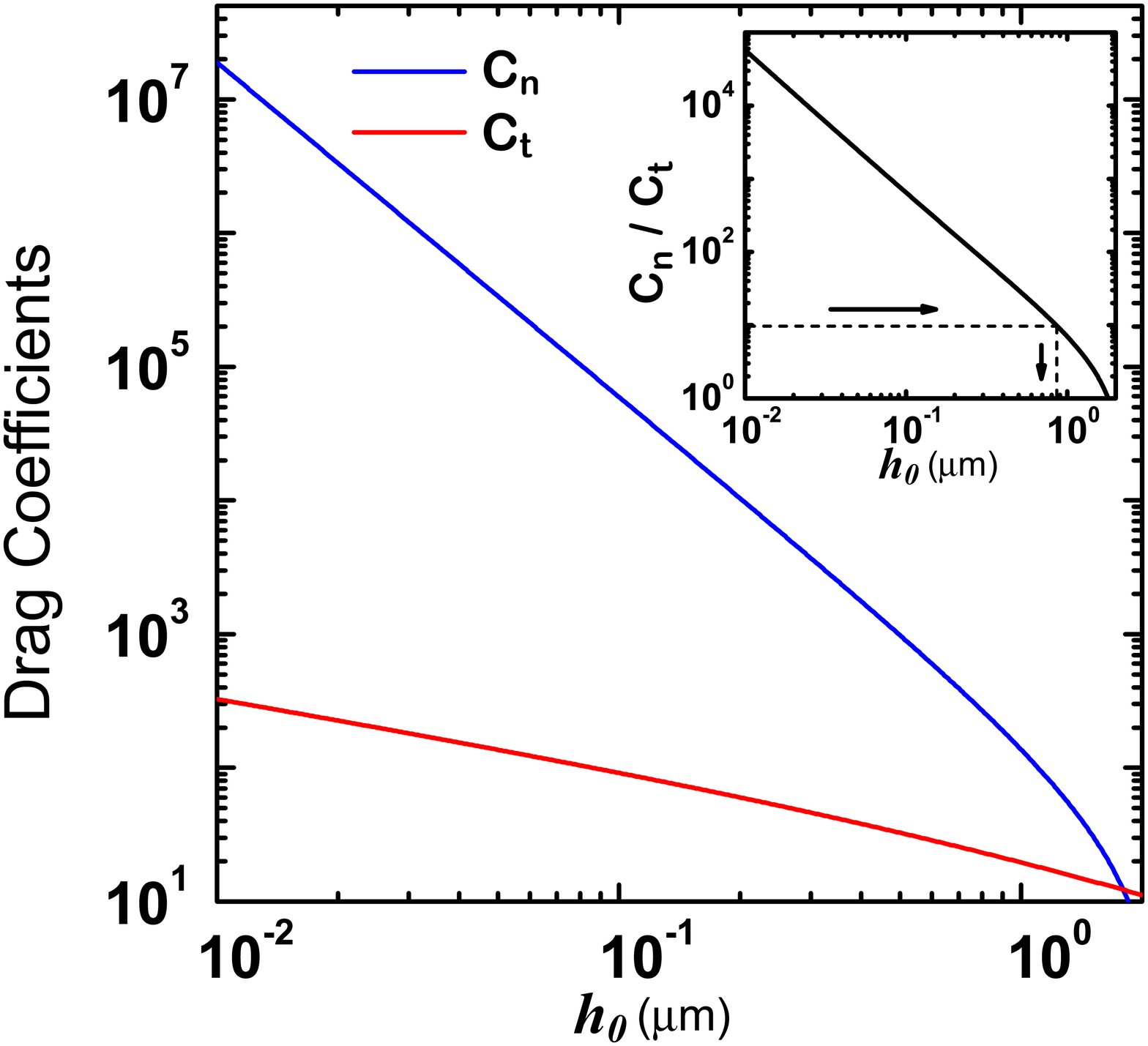}
      \caption{\label{DragCoef}(Color Online) The drag coefficients and the drag coefficient ratio (inset) as a function of the gap width $h_0$. The blue (upper) and red (lower) lines are the normal and tangential drag coefficients, respectively. Inset: The normal and tangential drag coefficient ratio $C_n/C_t$ as a function of $h_0$. The dashed lines show the estimated drag coefficient ratio $C_n/C_t \approx 9.4$ which corresponds to $h_0 \approx 0.85\mu$m.}
    \end{center}
\end{figure}

\subsection*{The effects of anisotropic drag forces on {\it C. elegans}' kinematic efficiency} Traditionally, two main motility gaits (swimming and crawling) are reported depending on whether {\it C. elegans} is fully immersed in a liquid or moving on top of a surface. Parameters used to distinguish between these two motility forms have included metrics such as bending frequency $f=1/T$, amplitude $A$, and wavelength $\lambda$, and values of such quantities are found to be much lower for crawling than for swimming~\cite{Fangyen,Pierce}. Our kinematic data also show similar trends (Table 1), and we find that the kinematic efficiency defined as $U/c$ is eight times larger for crawling than for swimming.

There are, however, some interesting similarities between crawling and swimming nematodes. In particular, the shape parameter value defined as $\lambda/A$ is nearly identical for both crawling and swimming gaits~(Table~1). This constant value of $\lambda/A$ is likely to result from biological constraints on the muscle strain~\cite{Karbowski}. Since $U/c$ is a function of $C_{n}/C_{t}$ and $\lambda/A$ only (see Eq.~\ref{EfficiencyEq}), we infer that the kinematic differences observed between these two motility gaits may be attributed to the different values of the ratio of drag coefficients experienced by {\it C. elegans} since $\lambda/A$ remains nearly constant.
\begin{table}%
\caption{\label{KineData}Kinematics for swimming and crawling nematode.}
     \setlength{\tabcolsep}{20pt}
     \begin{tabular}{lcc}
     \hline
     & Swimming & Crawling \\
     \hline
     Speed $U$ (mm/s)             & 0.36 $\pm$ 0.01 & 0.26 $\pm$ 0.01 \\
     Amplitude $A$ (mm)         & 0.26 $\pm$ 0.01 & 0.10 $\pm$ 0.01 \\
     Wavelength $\lambda$ (mm)  & 2.12 $\pm$ 0.08 & 0.79 $\pm$ 0.02 \\
     wavespeed  $c$ (mm/s)        & 4.20 $\pm$ 0.17 & 0.38 $\pm$ 0.02 \\
     \hline
     Shape parameter ($\lambda/A)$ & 8.14$\pm$ 0.29 & 7.84 $\pm$ 0.33 \\
     Efficiency ($U/c$)      &  0.08$\pm$ 0.01 & 0.67 $\pm$ 0.01 \\
     \hline
     \end{tabular}
\end{table}

Overall, we find that anisotropic drag forces influence the undulatory locomotion in two aspects, i.e. the energy cost (Eqs.~\ref{DragForceN} and~\ref{DragForceT}) and the undulation efficiency (Eqs.~\ref{EfficiencyEq}). The tangential drag coefficient $C_t$ determines the nematode's required propulsion (mechanical) force and the energy cost because it is responsible for most of the translational resistance to forward motion, while the drag coefficient ratio $C_n/C_t$ determines the ratio of the propulsion energy dissipated along both the transversal and longitudinal directions~\cite{Hancock}.

\begin{figure}
    \centering
    \includegraphics[width=0.85\textwidth]{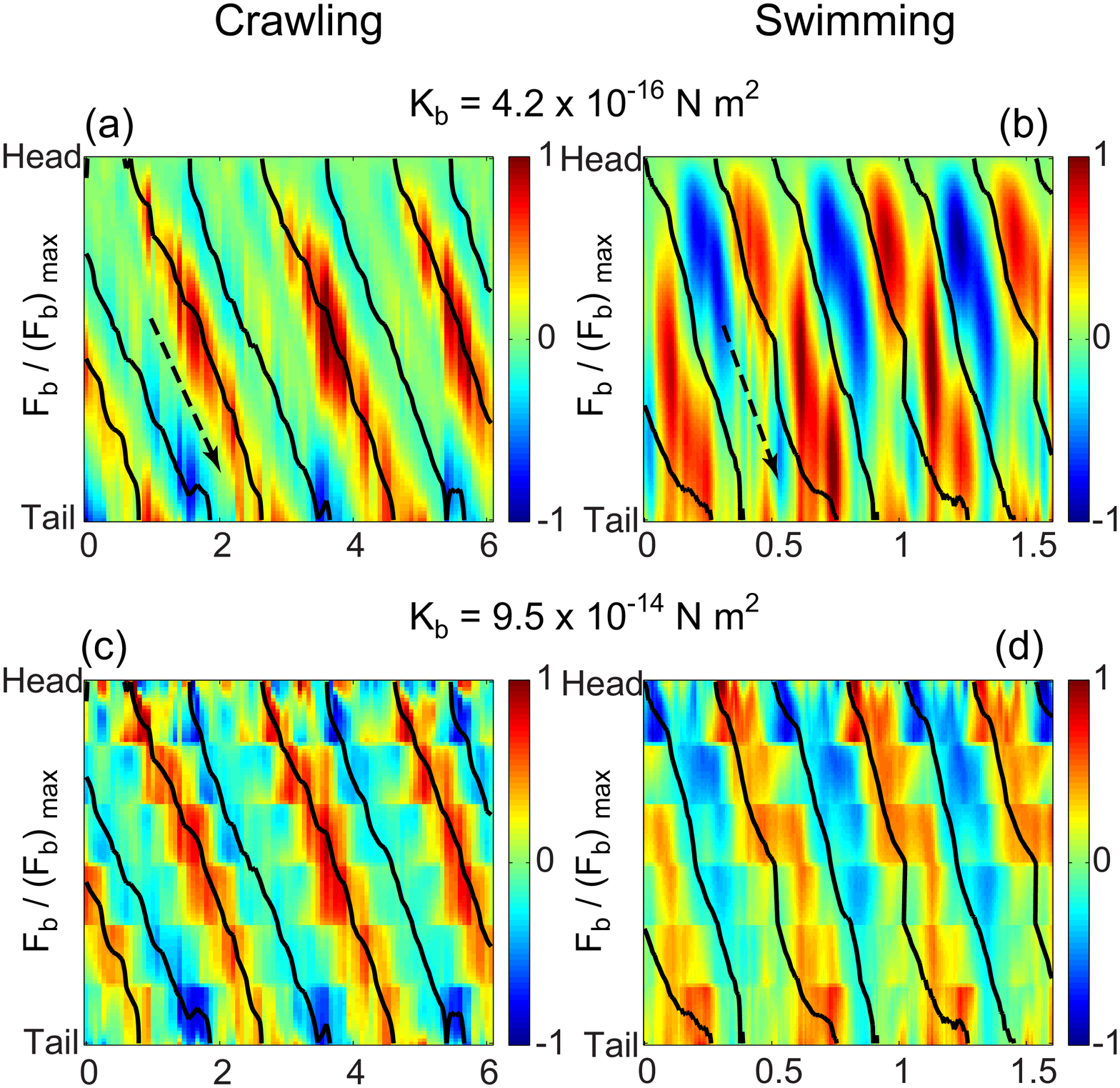}
    \caption{\label{SwimCrawl}(Color online) The normalized spatio-temporal bending force patterns for crawling (left-panel) and swimming (right-panel) {\it C. elegans}. The bending force calculated for (a,b) $K_b = 4.19\times 10^{-16}$ N m$^2$ \cite{SznitmanBioJournal} and (c, d) $K_b = 9.5 \times 10^{-14}$ N m$^2$ \cite{Fangyen}. The overlaid black solid lines corresponds to regions where the curvature $\kappa = 0$. The dashed arrow shows the direction of the traveling wave.}
\end{figure}

\subsection*{Calculating the nematode's internal bending force}
Bending is required for {\it C. elegans} to generate forward thrust. In this section, we calculate the total force necessary to bend the nematode's body using experimental data and knowledge of the drag forces. We compare the calculated nematode bending force patterns between crawling and swimming nematodes. A force balance on the nematode body shows that the bending force needed to overcome resistance arises from two components: one is the elastic resistance due to body bending, while the other is the viscous force distributed along the nematode body. In the limit of small undulating amplitudes, the total muscle bending force $\vec{F}$ at position $s = s_p$ can be obtained by summing up the elastic and distributed viscous drag forces acting along the nematode body,
\begin{flalign}
\vec{F}(s_p, t) = K_b \frac {\partial \kappa(s_p, t)}{\partial s} \vec{n}(s_p,t) + \mu \int_0^{s_p} C_n \vec{u}_n(s, t) + C_t \vec{u}_t(s,t)\text{ }ds,
\label{eq:totalforce}
\end{flalign}
where $s$ = $0$ and $s$ = $L$ represent the head and tail of the nematode body, respectively. The unit normal vector of the local body segment is given by $\vec{n}(s_p, t)$. The first term on the right-hand side represents the elastic force required to bend the nematode body, while the second term represents the viscous force due to the liquid film. Altogether, internal muscle actuation must overcome both the elastic and viscous drag forces to sustain undulatory locomotion (see~\cite{SM} for details).

The total muscle force $\vec{F}$ is decomposed into normal and tangential directions. The tangential component of the force describes the resistance to changes in body length. The normal component of $\vec{F}(s_p, t)$ describes the bending force, and it is given by

\begin{flalign}
  F_b(s_p, t) &= \vec{F}(s_p, t) \cdot \vec{n}(s_p, t).
\end{flalign}

The magnitude of the internal bending force $F_b(s,t)$ can be determined at any time $t$ and location $s$ along the nematode body by combining experimentally measured nematode speed and bending curvature with individual values of the normal and tangential drag coefficients. To compute the internal elastic contribution, we also need values of the nematode's bending modulus $K_b$, a material property that is typically difficult to measure. Here, we will use two previously estimated values of $K_b$, namely $K_b = 4.2 \times 10^{-16}$ N m$^2$~\cite{SznitmanBioJournal} and $K_b= 9.5 \times 10^{-14}$ N m$^2$ ~\cite{Fangyen}. The differences in the magnitude of $K_b$ reflect the different assumptions in determining the spatial and temporal form of the nematode's (internal) active bending moment. Figure~\ref{SwimCrawl}(a, b) shows the magnitude of $F_b$ using $K_b = 4.2 \times 10^{-16}$ N m$^2$~\cite{SznitmanBioJournal} for both crawling and swimming nematodes for three bending cycles. Figure~\ref{SwimCrawl}(c, d) shows the corresponding magnitude of $F_b$ using $K_b= 9.5 \times 10^{-14}$ N m$^2$~\cite{Fangyen}. The solid black lines in Fig.~\ref{SwimCrawl} correspond to regions where the nematode's curvature $\kappa$ is zero, and arrows show the direction of the bending wave.

The bending force patterns presented in Fig.~\ref{SwimCrawl} are time-periodic and in general, spatially complex. For swimming nematodes with $K_b = 4.2 \times 10^{-16}$ N m$^2$~\cite{SznitmanBioJournal}, the bending force is larger near the head, while most of the bending force is accumulated around the mid-section for crawling {\it C. elegans}. For both swimming and crawling {\it C. elegans}, spatial patterns for the bending force are very similar to previously described gait-specific features of calcium signals in muscle~\cite{Pierce}. In particular, we note that the larger value for $K_b$ (Fig.~\ref{SwimCrawl}c,d) yields larger resistance to nematode bending. As a result, the spatial distribution of $F_b$ is not identical for the two cases.

\begin{figure}
    \centering
    \includegraphics[width=0.85\textwidth]{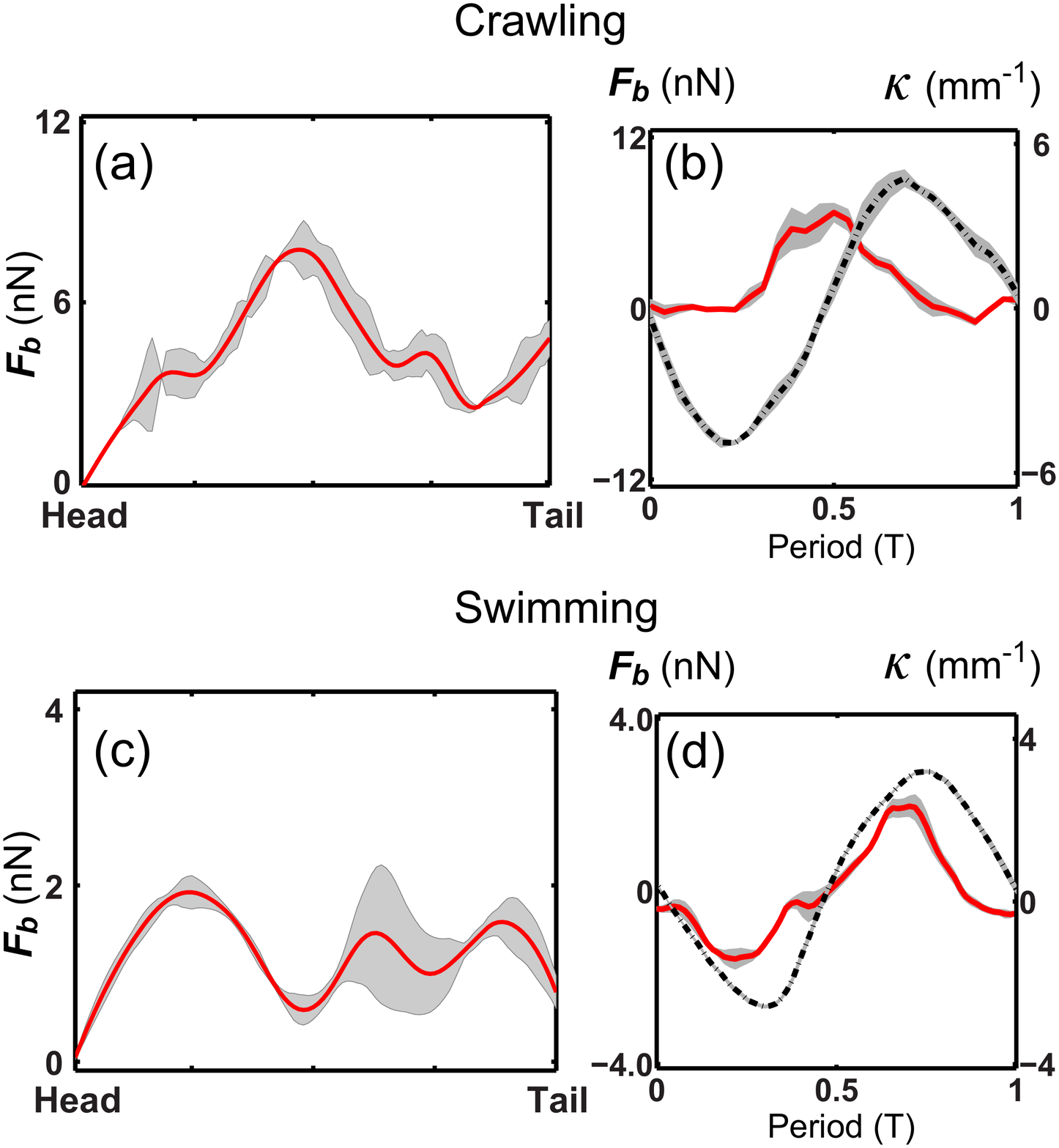}
    \caption{\label{ForceTempo} Spatial (a,c) and temporal (b,d) patterns of the bending forces $F_b$ for crawling (top-panel) and swimming (bottom-panel) {\it C. elegans} for $K_b = 4.19\times 10^{-16}$ N m$^2$. The shaded areas correspond to the standard error of the mean (SEM) (a,c) The spatial patterns are averaged over five periods. Most of the bending force is located near the head for swimming nematodes and around the body's mid-section for crawling. (b,d) Temporal patterns are obtained at position $s/L=0.55$ for both crawling and swimming nematodes, and the dashed line corresponds to the bending curvature.}
\end{figure}

Since the bending force patterns are time-periodic, the data shown Fig.~\ref{SwimCrawl}(a,b) can be phase-averaged. Here, we focus our analysis on the $K_b =  4.2\times 10^{-16}$ N m$^2$~\cite{SznitmanBioJournal} case only. Figure~\ref{ForceTempo}(a) and (c) shows the phase-averaged spatial distribution of $F_b$ for crawling and swimming nematodes, respectively. Note that the shaded areas in Fig.~\ref{ForceTempo} correspond to standard error of the mean (SEM). In general, crawling nematodes generate bending forces $F_b$ that are fourfold larger than nematodes swimming in an aqueous solution. The peak values of internal bending force for crawling and swimming nematodes are approximately 8.5 nN and 2.0 nN, respectively. As noted before, the bending force during crawling appears concentrated in the middle section of the body.

The phase-averaged bending force signal is shown as a function of one bending period in Fig.~\ref{ForceTempo}(b) and (d) for crawling and swimming nematodes. The bending force signals are compared to the curvature signals, and measurements are made at $s/L=0.55$ for both swimming and crawling. Results are shown for $K_b = 4.2\times 10^{-16}$ N m$^2$ only. We find that, for crawling, the bending force and bending curvature signals are not in phase with each other (Fig.~\ref{ForceTempo}b). On the other hand, the force and curvature signals are nearly in phase for swimming nematodes. One possible explanation for the observed phase lag in crawling nematodes is the (much) larger viscous drag forces experienced by nematodes moving on wet surfaces. In the case where little viscous drag is present, the internal bending forces remain nearly in phase with the bending curvatures, as indicated by Eq. 14. In other words, a phase differences between the bending force and curvature is a consequence of enhanced viscous drag forces; the larger the drag forces, the larger the phase difference between the internal bending forces and the bending curvatures. Finally, we note that for both gaits, the bending curvature signals are rather periodic, while the bending force signals exhibit unequal strength along ventral and dorsal sides. This unequal ventral/dorsal bending is most likely the result of an asymmetry of the ventral/dorsal neurological structures~\cite{White} (see Fig. 4).

\begin{figure}
  \centering
  \includegraphics[width=0.85\textwidth]{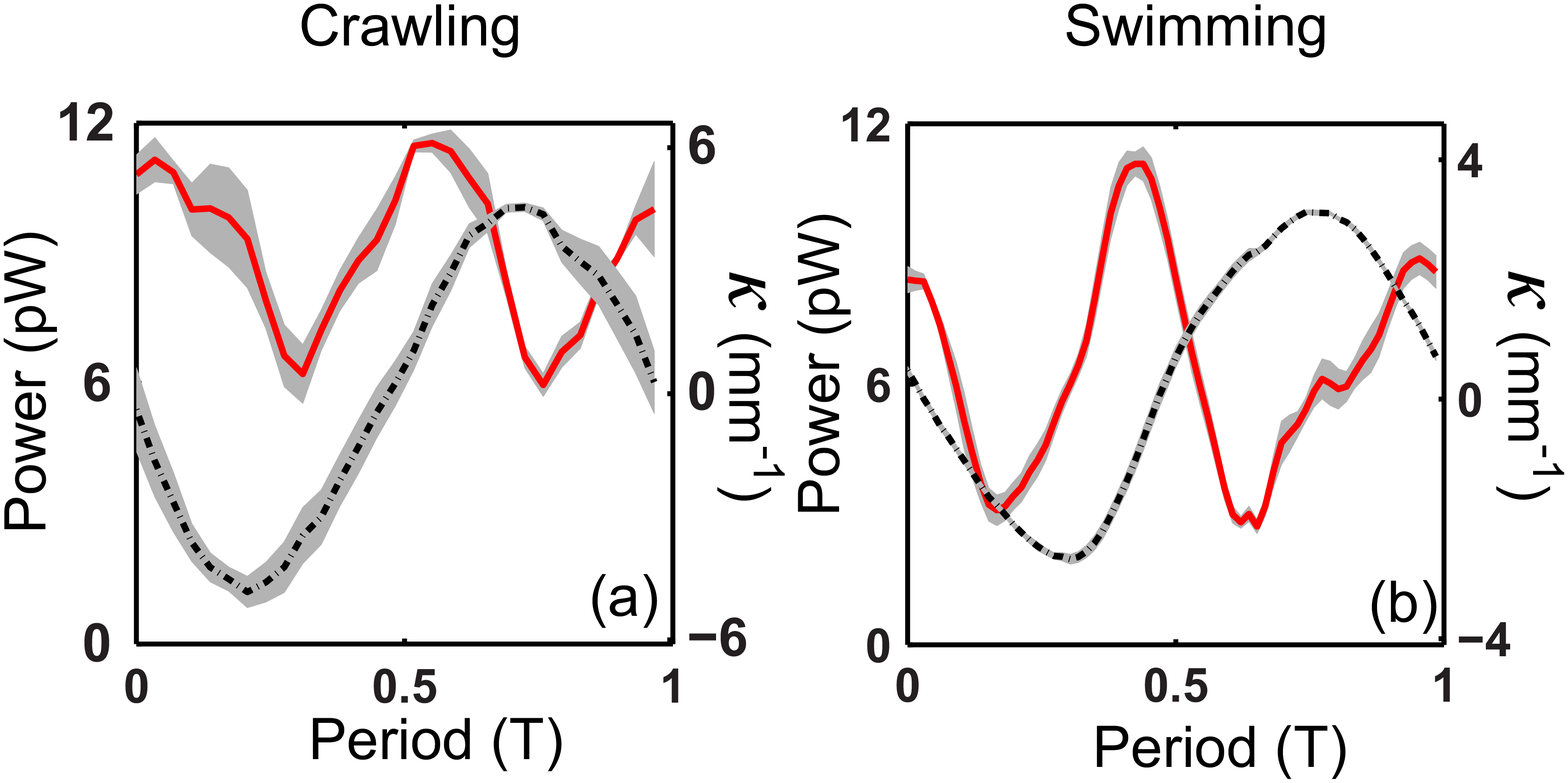}\\
  \caption{(Color online) Mechanical power dissipated during (a) crawling and (b) swimming. The solid red and dashed black curves are the dissipated mechanical power and bending curvatures. The shaded areas shown the standard error of the mean (SEM). Nematodes dissipate more energy during crawling compared to swimming.}\label{Power}
\end{figure}

We can further study the biomechanics of undulatory locomotion by computing the power dissipated by the nematodes during crawling and swimming. As shown above, the internal bending force is composed by the elastic and viscous forces, and the mechanical power dissipated by {\it C. elegans} during locomotion can be computed as

\begin{equation}
  P(t) = \mu \int_0^L C_n u_n^2(s, t) + C_t u_t^2(s,t)\text{ }ds.
\end{equation}
Since the elastic energy is conserved, energy dissipation is then due to the viscous forces. In Fig.~\ref{Power}, we show the dissipated power for both crawling and swimming nematodes over one beating cycle, respectively. We find that the crawling nematodes dissipate more energy than swimming nematodes.  The time-averaged power is  6.4 pW and 4.5 pW for crawling and swimming nematodes, respectively. Although the magnitudes in power are different, both the swimming and crawling gaits display similar temporal patterns. The phase differences between the power and the bending curvature temporal signals are also quite similar for both motility gaits.

\section*{Discussion}
Uncoordination ({\it unc}) is a very common phenotype among genetic mutations in {\it C. elegans}. The affected genes may be expressed either in neurons, muscles or even in other tissues affecting the structures required for coordinated movement~\cite{unc}. For more than 30 years, different {\it unc} phenotypes of {\it C. elegans} were mostly observed on NGM (agar) plates. Here, we have quantified the biomechanical properties of crawling {\it C. elegans} crawling using both experiment and modeling in an effort to provide quantitative tools for motility phenotyping of mutant strains on wet agar gels. The analysis presented here focuses on wild-type {\it C. elegans}, although the methods can be easily extended to different nematode's strains and other organisms. We coupled kinematic data, such as nematode's speed and body curvature, with a hydrodynamic model in order to obtain estimates of the (surface) drag forces acting on the {\it C. elegans}. Values of the surface drag forces were then used to calculate the nematode's internal (muscle) bending force, which are compared to recent calcium imaging data~\cite{Pierce}. Results on crawling nematodes are also compared to {\it C. elegans} swimming in water-like, buffer solutions in order to gain insight into the nematode's adaptive motility kinematics across different environments.

Results show that the expressions for the surface drag forces along the normal and tangential directions of the nematode's body (Eqs.~\ref{DragForceN},\ref{DragForceT}) share a similar form with drag forces obtained for swimming using resistive force theory (RFT). Equations~\ref{DragForceN} and \ref{DragForceT} indicate that resistance to locomotion can be quantified by the tangential drag coefficient $C_t$ while propulsion efficiency is a strong function of the drag coefficient ratio $C_n/C_t$. We find that the drag coefficient ratio $C_n/C_t$ is approximately 10 for nematodes moving on wet agar plates, in agreement with a recent study~\cite{Karbowski}. Other investigations estimate values of $C_n/C_t$ that range from less than 2~\cite{Sauvage} to larger than 20~\cite{Cohen} for crawling nematodes. For comparison, the values of $C_n/C_t$ range from 1.4 to 2 for swimming nematodes. By considering a thin liquid film between the nematode's body and the agar surface, we are also able to estimate the individual values of the normal and tangential components of the drag forces for crawling nematodes. Using Eqs. \ref{DragForceN} and \ref{DragForceT} along with kinematic data, we find that $C_n=222.0$ and $C_t=22.1$ (Fig.~\ref{DragCoef}). For comparison, we note that a recent theoretical analysis estimated $C_n$ and $C_t$ to be of $O(10)$, with a drag coefficient ratio less than 2~\cite{Sauvage}, which is closer to swimming.

In addition, our analysis of the drag forces suggests that gait modulation (crawling versus swimming) is probably a response to the external environment, or more specifically, to the drag coefficient ratio ($C_n/C_t$). That is, {\it C. elegans} seems to adjust not only the magnitude of the internal bending force, but more importantly its distribution along the body, to accommodate for changes in $C_n/C_t$ and maintain its propulsive thrust, as shown in Fig.~\ref{SwimCrawl}. A recent investigation~\cite{Fangyen} has shown that the power required for a nematode to swim in highly viscous fluids is orders of magnitude larger (5 nW) than found here for crawling, yet the ratio $C_n/C_t$ for swimming remains constant at approximately $2$ and is independent of fluid viscosity. Hence, nematodes maintain their distinct swimming gait even if the forces required to locomote are much larger than during crawling. Only when the drag coefficient ratio $C_n/C_t$ is modified is that nematodes exhibit crawling-like gaits or behavior. Our present results suggest that {\it C. elegans}' motility gait can be tuned and modulated by modifying the anisotropy of the surface drag coefficients.

Knowledge of the individual components of the drag forces allows for calculating the internal bending force of both crawling and swimming nematodes (Fig.~\ref{SwimCrawl}). We find that the bending force patterns are time-periodic and spatially complex. These patterns provide additional insight into the different motility gaits observed in {\it C. elegans}, namely swimming and crawling. The temporal patterns of the bending force, for example, reveal a phase lag with respect to the bending curvature. This phase lag is a measure of the viscous drag force contribution, which is larger for crawling than for nematodes swimming in water-like buffer solutions. The spatial patterns show that large values of the bending force are concentrated near the head for swimming nematodes, while, for crawling, much of the bending force is concentrated in the nematode's mid-section. These calculated spatial patterns correlate rather well with regions of high calcium activity~\cite{Pierce}. The analysis and data presented here strongly suggest that while the motility patterns associated with crawling and swimming can be distinguished by distinct kinematics (Fig.~\ref{DragCoef} and Table 1) and bending force patterns (Fig.~\ref{SwimCrawl}), changes in the environment's (external) drag coefficient ratio can also contribute to modulations in motility gaits (Eqs. 11 and 12). Further, the analysis is able to capture main biomechanical features of undulatory crawling on lubricated surfaces (agar plates), and it can provide direct estimates of force and power expenditures that are useful for motility phenotyping activities such as in genetic and drug screening applications involving {\it C. elegans}.

\section*{Acknowledgments}
We acknowledge fruitful discussions with S. Jung, A. Brown, R. Carpick, G. Wabiszewski, G. Juarez, and N. Keim. This work was supported by NSF-CAREER (CBET)-0954084.


\end{document}